\documentclass[aps,prb,twocolumn,floatfix,footinbib,superscriptaddress]{revtex4}
\usepackage[english]{babel}
\usepackage[utf8]{inputenc}
\usepackage[dvips]{graphicx}
\usepackage{fancyhdr}
\usepackage[active]{srcltx}
\usepackage{setspace}
\usepackage{float}
\usepackage{amsmath}
\usepackage{amsfonts}
\usepackage[]{natbib}
\usepackage{array}
\usepackage{color}
\usepackage{multirow}

\begin{document}

\title{$k \cdot p$ theory for phosphorene: Effective g-factors, Landau levels, and excitons}

\author{Paulo E. Faria~Junior}
\email{fariajunior.pe@gmail.com}
\affiliation{Institute for Theoretical Physics, University of Regensburg, 93040 Regensburg, Germany}

\author{Marcin Kurpas}
\affiliation{Institute of Physics, University of Silesia, 75 Pułku Piechoty 1, 41-500 Chorzów, Poland}

\author{Martin Gmitra}
\affiliation{Institute of Physics, P. J. Šafárik University in Košice, Park Angelinum 9, 04001 Košice, Slovakia}

\author{Jaroslav Fabian}
\affiliation{Institute for Theoretical Physics, University of Regensburg, 93040 Regensburg, Germany}


\begin{abstract}

Phosphorene, a single layer of black phosphorus, is a direct-band gap 
two-dimensional semiconductor with promising charge and spin transport properties.
The electronic band structure of phosphorene is strongly affected by the structural 
anisotropy of the underlying crystal lattice. We describe the relevant conduction 
and valence bands close to the $\Gamma$ point by four- and six-band (with spin) 
$k \cdot p$ models, including the previously overlooked interband spin-orbit coupling 
which is essential for studying anisotropic crystals. All the $k \cdot p$ parameters 
are obtained by a robust fit to {\it ab initio} data, by taking into account the 
nominal band structure and the $k$-dependence of the effective mass close to $\Gamma$-point. 
The inclusion of interband spin-orbit coupling allows us to determine dipole transitions 
along both armchair and zigzag directions. The interband coupling is also key to 
determine the effective g-factors and Zeeman splittings of the Landau levels. 
We predict the electron and hole g-factor correction of $\approx 0.03$ due to the intrinsic 
contributions in phosphorene, which lies within the existing range of experimental 
data. Furthermore, we investigate excitonic effects using the $k \cdot p$ models 
and find exciton binding energy (0.81 eV) and exciton diameters consistent with 
experiments and {\it ab initio} based calculations. The proposed  $k \cdot p$ Hamiltonians 
should be useful for investigating magnetic, spin, transport, optical properties and 
many-body effects in phosphorene. 

\end{abstract}


\maketitle


\section{Introduction}
\label{sec:intro}

The two-dimensional (2D) phosphorene was first synthesized in 2014\cite{Liu2014ACSNano, 
Lu2014NR, Liang2014NL, Zhang2014ACSNano, CastellanosGomez20142DMat, Qiao2014NatComm}
and showed remarkable physical properties. For example, phosphorene is a direct 
band gap semiconductor with enhanced photoluminescense intensity compared to bulk 
black phosphorus\cite{Zhang2014ACSNano}. Moreover, its strong coupling to light can be 
varied  in the far-infrared to red spectral range\cite{Zhang2014ACSNano} due to 
high sensitivity of the band gap to the number of monolayers\cite{Keyes1953, Das2014NL, 
Liang2014NL, CastellanosGomez20142DMat, Wang2015NatNano, Frank2019PRX}.
Due to the puckered crystalline structure phsophorene shows strongly anisotropic 
electronic properties\cite{Liu2014ACSNano,Qiao2014NatComm}. High carrier mobility 
in phosphorene allows not only observation of fundamental quantum phenomena, such 
as Landau levels\cite{Li2015NatNano, Gillgren20152DMat} or quantum Hall effect\cite{Li2016NatNano}, 
but also its potential applications to semiconductor spintronics\cite{Zutic2004RMP,Han2014NatNano}, 
due to the weak spin-orbit coupling (SOC) of phosphorus\cite{Kurpas2016PRB}. Indeed, nanosecond 
spin lifetimes observed in all-electrical spin injection experiments and realization 
of a spin valve operating at room temperature\cite{Avsar2017NatPhys} already 
demonstrated the robustness of spin coherence in phosphorene. Furthermore, few-layer 
phosphorene heterostructures are promising candidates for ultrafast switching based 
on optically generated surface polaritons\cite{Huber2017NatNano}.

There have already been applications of the $k \cdot p$ method to describe 
phosphorene.\cite{Rodin2014PRL,Li2014PRB,LewYanVoon2015NJP,PereiraJr2015PRB,Kafaei2018JAP}
However, to also study spin-orbit effects the effective models should fully exploit 
the symmetry of phosphorene by capturing the anisotropy of the interband dipole coupling. 
Effective g-factors illustrate this best. For monolayer transition metal dichalcogenides (TMDCs) 
the g-factors can be derived employing the conventional $\vec{p}$ matrix elements\cite{Kormanyos2013PRB,Kormanyos2014PRX,Kormanyos2015NJP} 
(equivalent to the so called Kane matrix element in zinc-blende structures\cite{Kane1957JPCS}), 
in the perturbative fashion within the $k \cdot p$ framework\cite{Roth1959PR,Hermann1977PRB,LewYanVoon2009}.
In phosphorene, due to its two-fold symmetry embedded in the $D_{2h}$ group, the usual Kane-like 
matrix elements can account only for the coupling along the armchair direction. In order 
to include the contribution along the zigzag direction, one needs to go beyond the conventional 
$\vec{p}$ matrix element and include the $k$-dependent SOC contribution. The inclusion 
of this term was already shown by Zhou et al.~[\onlinecite{Zhou2017PRB_gph}] to 
provide the g-factor correction in phosphorene thin films with its value estimated 
from experimental data. As another example, the inclusion of such $k$-dependent 
SOC terms in III-V wurtzite semiconductors was recently shown to provide a more 
reliable fitting to the {\it ab initio} spin splitting and overall band structure\cite{FariaJunior2016PRB} 
and also to add sizable corrections to the total value of the effective g-factors\cite{FariaJunior2019PRB}. 
Furthermore, more complete $k \cdot p$ models for phosphorene can be used as building 
blocks to model van der Waals heterostructures\cite{Geim2013Nat}, for instance combined 
with TMDCs\cite{Kormanyos20152DMat} and (In,Ga)Se materials\cite{Zhou2017PRB_kpInSe}, 
overcoming computational costs of {\it ab initio} calculations.

In this work, we investigate important physical features which appear due to the 
inclusion of the interband SOC term in effective $k \cdot p$ models for monolayer 
phosphorene. We show that this additional SOC term not only provides the interband 
dipole coupling along zigzag direction but also allows us to predict the values 
of the effective g-factors from a full theoretical perspective highlighting the 
intrinsic contributions of monolayer phosphorene. We also analyze the Landau 
level (LL) spectra and show that the interband SOC term provides a correction 
to Zeeman splitting, consistent with the g-factor approach. Finally, we combine 
our effective models with the Bethe-Salpeter equation (BSE) and show that the excitonic 
spectra of monolayer phosphorene are in  agreement with the available data in the 
literature. We point out that our $k \cdot p$ parameters are obtained from a systematic 
fitting to {\it ab initio} band structure calculations taking into account the 
$k$-dependence of the energy bands and the effective masses (weighting the contribution around $\Gamma$-point), 
a crucial point to correctly describe the linear-in-k couplings in the Hamiltonian. 
Besides providing a simplified and tangible understanding of the underlying physics, 
these effective $k \cdot p$ models might be used to investigate additional properties 
in phosphorene but also of more complex systems such as van der Waals heterostructures 
composed of several layered materials.

The paper is organized as follows: In Sec.~\ref{sec:abinitio} we discuss the 
{\it ab initio} band structure of phosphorene. The effective $k \cdot p$ models with the inclusion 
of the interband SOC are addressed in Sec.~\ref{sec:kp}. In Sec.~\ref{sec:bmag}
we investigate the behavior of the different $k \cdot p$ models under external magnetic field 
by calculating the effective g-factors and the LL spectra. Excitonic effects 
are presented in Sec.~\ref{sec:excitons} and finally, in Sec.~\ref{sec:conclusions}
we draw our conclusions. 


\section{Phosphorene band structure from {\it ab initio}}
\label{sec:abinitio}

The initial crystal structure of phosphorene layer was taken from the bulk black 
phosphorus\cite{Brown1965}. New cell parameters were then found by fully relaxing 
a sheet of phosphorene using  quasi-Newton variable-cell scheme as implemented 
in  the {\sc Quantum Espresso} package\cite{QE2009,QE2017}. During this process all 
atoms were free to move in all directions in order to relax internal forces. The 
force convergence threshold and total energy convergence threshold were set to $10^{-4}$~Ry/a.u. 
and $10^{-5}$~Ry/a.u., respectively. A vacuum of 20$~\rm \AA$  was introduced in order 
to reduce spurious interactions between the periodic copies of the system. We used the norm-conserving 
pseudopotential with the Perdew-Burke-Ernzerhof (PBE)\cite{perdew_1996} version 
of the generalized gradient approximation (GGA) exchange-correlation potentials, 
with kinetic energy cutoffs of 70 Ry and 280 Ry for the wavefunction and charge 
density, respectively. The optimized cell parameters are a = 3.2986~\AA~ along 
the zigzag edge, and b = 4.6201~\AA~along the armchair edge. The electronic and 
spin properties of phosphorene were calculated using the full-potential augmented 
plane-wave all-electron code package WIEN2k\cite{wien2k}. Self-consistency was 
achieved with 151 k-points in the irreducible wedge of the Brillouin zone. Tuning 
of the band gap was achieved by combining the undressed LDA functional with the 
modified Becke-Johnson potential\cite{mBJ}. The parameters of the latter were 
chosen to give the gap close to the recent experimental\cite{Liang2014NL,Wang2015NatNano} 
and theoretical\cite{Frank2019PRX} values and to provide a realistic description 
of phosphorene.

\begin{figure}[h!]
\begin{center} 
\includegraphics{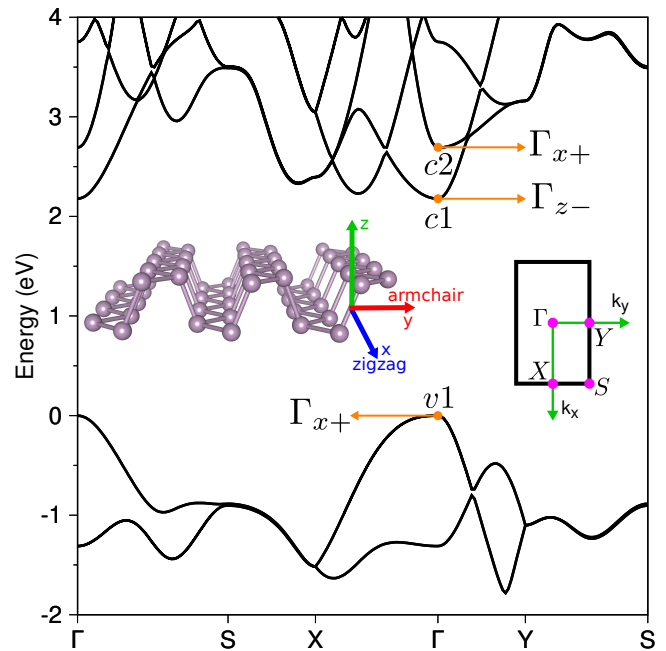}
\caption{(Color online) {\it Ab initio} band structure of phosphorene. The labels 
at $\Gamma$-point indicate the energy bands used to construct the effective $k \cdot p$ 
Hamiltonians, i. e., first valence ($v1$), first conduction ($c1$) and second conduction 
($c2$) bands with the irreducible representations from the single group $D_{2h}$. 
In the inset we show the crystal structure of phosphorene with the identified armchair and zigzag directions, and the first Brillouin 
zone highlighting the high-symmetry points and momentum directions.}
\label{fig:abinitio}
\end{center}
\end{figure}

The resulting {\it ab initio} band structure is shown in Fig.~\ref{fig:abinitio}. 
The calculation reproduces the direct band gap of phosphorene with the value of
2.178~eV centered at the $\Gamma$ point. 
This value is close to the recent quantum Monte Carlo result of 2.4 eV\cite{Frank2019PRX}.
The energy dispersion of the top-most 
valence (labeled $v1$) and bottom-most conduction (labeled $c1$) bands close to 
the band edge displays a sizeable anisotropy with respect to the main crystallographic 
directions. This anisotropy is particularly pronounced for the band $v1$, which 
is almost flat for momenta along the $\Gamma$-X direction (along the zigzag edge 
of phosphorene), and very dispersive along the $\Gamma$-Y path (along the armchair 
edge do phosphorene). For this band, the corresponding effective masses $m_{v1,x}$ 
and $m_{v1,y}$ differ by more than an order of magnitude (see Section \ref{sec:kp}). 

Phosphorus is a light element, therefore the influence of spin-orbit coupling on 
the band structure is mainly limited to the removal of orbital degeneracies at certain 
high-symmetry points and lines, e.g., at the X-S and Y-S paths. At the S point, 
the splitting is 21~meV and 17~meV for the valence band $v1$ and conduction band 
$c1$, respectively\cite{Kurpas2016PRB}. The spin degeneracy of the energy bands 
is not removed by SOC, by virtue of space inversion and time reversal symmetry. 
We also identify in Fig.~\ref{fig:abinitio} the single group irreducible representations (irreps) 
for the energy bands at $\Gamma$ point that are used as input to the k.p Hamiltonians 
discussed in the next section.

\section{$k \cdot p$ modeling}
\label{sec:kp}

For the effective description of the phosphorene band structure, given in Fig.~\ref{fig:abinitio}, 
we focus on the first valence ($v1$), first conduction ($c1$) and second conduction 
($c2$) bands, identified by the irreps of the symmetry group $D_{2h}$ provided by 
the {\it ab initio} calculations. This irrep identification of the energy bands is a crucial 
step for the development of $k \cdot p$ Hamiltonians\cite{Dresselhaus1955PR,Dresselhaus2008,FariaJunior2016PRB}. 
The Hamiltonian basis set, including spin, is then given by
$\left\{ \left|\Gamma_{x+}^{c2}\uparrow\right\rangle ,\left|\Gamma_{x+}^{c2}\downarrow\right\rangle ,\left|\Gamma_{z-}^{c1}\uparrow\right\rangle ,\left|\Gamma_{z-}^{c1}\downarrow\right\rangle ,\left|\Gamma_{x+}^{v1}\uparrow\right\rangle ,\left|\Gamma_{x+}^{v1}\downarrow\right\rangle  \right\}$, 
in which the orbital part $\left|\Gamma_{\alpha}^{n}\right\rangle$ is written in Dirac 
notation, defined as $\left\langle \vec{r} | \Gamma_{\alpha}^{n}\right\rangle = u_\alpha^n(\vec{r})$. 
Essentially, our notation means that the Bloch function $u_\alpha^n(\vec{r})$ of the energy 
band $n$ transforms as the irrep $\Gamma_\alpha$ of the symmetry group $D_{2h}$. The 
vertical arrows represent the spin up and down projections, eigenvalues of the $\sigma_z$ 
Pauli matrix. In Sec.~I of the Supplemental Material\cite{supmat} we summarize 
the symmetry properties of the group $D_{2h}$ of phosphorene.

Within the $k \cdot p$ framework, we must compute matrix elements between different energy 
bands that are mediated by vector (such as $\vec{p}$) or pseudovector (such as 
$\vec{\nabla} V \times \vec{p}$) operators. In the language of group theory, we 
would write these matrix elements as the direct product $\Gamma^n \otimes \Gamma^o \otimes \Gamma^m$, 
with $\Gamma^{n(m)}$ representing the irrep of the energy band $n(m)$ and $\Gamma^o$ 
representing the irrep of the operator. If the result of such direct product contains 
the identity irrep (with characters equal to 1 for all symmetry operations), then 
the matrix element is nonzero\cite{Dresselhaus2008}. Furthermore, since the irreps 
of the symmetry group $D_{2h}$ are all one dimensional, the nonzero matrix elements 
are readily available by inspecting the multiplication table, given in Sec.~I 
of the Supplemental Material\cite{supmat}. We point out that a thorough and systematic investigation 
of the symmetry properties of phosphorene has been performed by Li and Appelbaum\cite{Li2014PRB}, 
however without providing realistic values for the $k \cdot p$ parameters. 
In our study, we go beyond the derivation of $k \cdot p$ Hamiltonians based on symmetry arguments 
and also focus on the estimation of the Hamiltonian parameters by using the {\it ab initio} 
data presented in Sec.~\ref{sec:abinitio}.

Our strategy is to investigate different $k \cdot p$ formulations, including different energy 
bands and couplings, to demonstrate qualitative differences stemming from various couplings. In what follows we define the different $k \cdot p$ models we considered: 
ph6 (describes $c2$, $c1$, and $v1$ bands), ph4 (describes $c1$ and $v1$ bands), 
and Nph4 (also describes $c1$ and $v1$ bands but without any coupling between them). 
For the most general model, ph6, we can write the total Hamiltonian identifying the different 
coupling blocks
\begin{equation}
H_{\text{ph6}}=\left[\begin{array}{ccc}
H_{c2} & H_{c2c1} & H_{c2v1}\\
H_{c2c1}^{\dagger} & H_{c1} & H_{c1v1}\\
H_{c2v1}^{\dagger} & H_{c1v1}^{\dagger} & H_{v1}
\end{array}\right] \, ,
\label{eq:Hph6}
\end{equation}
with each coupling block being a $2 \times 2$ matrix. To obtain the Hamiltonian 
for the ph4 model, $H_{\text{ph4}}$, we just remove the blocks that have contribution 
of the band $c2$, i. e., the 1st row and the 1st column of Eq.~(\ref{eq:Hph6}). For 
the model Nph4, we can obtain the Hamiltonian $H_{\text{Nph4}}$ by just removing 
the interaction block $H_{c1v1}$ between $c1$ and $v1$ bands from $H_{\text{ph4}}$.

Each of the Hamiltonian blocks in Eq.~(\ref{eq:Hph6}) can be written as
\begin{equation}
H_{a}=H_{0,a}+H_{k,a}+H_{k2,a} \, ,
\end{equation}
with $H_{0,a}$ containing only $k$-independent terms, $H_{k,a}$ containing terms 
linear in $k$, $H_{k2,a}$ containing terms quadratic in $k$ with the subindex $a$ 
indicating the specific block. Specifically, taking into account the symmetry properties 
of phosphorene, the $k$-independent Hamiltonian block is given by
\begin{equation}
H_{0,a}=\left[\begin{array}{cc}
E_{a} & 0\\
0 & E_{a}
\end{array}\right]  \, ,
\end{equation}
which appears only for $a=\{c2,c1\}$ and the parameters $E_a$ indicate the 
energy values at $\Gamma$-point (note that $E_{c1}=E_g$). 
The Hamiltonian block with linear contribution of $k$ is given by
\begin{equation}
H_{k,a}=\left[\begin{array}{cc}
-iP_{a}k_{y}-\alpha_{a}k_{x} & 0\\
0 & -iP_{a}k_{y}+\alpha_{a}k_{x}
\end{array}\right]  \, ,
\end{equation}
which appears only for $a=\{c2c1,c1v1\}$ and the parameters $P_a$ and $\alpha_a$ 
originate from first-order perturbation theory of the operator
\begin{equation}
\vec{\Pi}=\frac{\hbar}{m_{0}}\vec{p}+\frac{\hbar^{2}}{4m_{0}^{2}c^{2}}\left[\vec{\sigma}\times\vec{\nabla}V(\vec{r})\right] \, ,
\label{eq:Pi}
\end{equation}
with $P_a$ being the conventional Kane-like matrix elements [originated from the 1st term in Eq.~(\ref{eq:Pi})] 
and $\alpha_a$ being the $k$-dependent interband SOC parameter [originated from the 2nd term in Eq.~(\ref{eq:Pi})]. 
We note that due to the symmetry of energy bands involved, the $\alpha_a$ terms do 
not mix different spins. Finally, the Hamiltonian block with quadratic contribution 
of $k$ is given by 
\begin{equation}
H_{k2,a}=\frac{\hbar^{2}}{2m_{0}}\left[\begin{array}{cc}
A_{a}k_{x}^{2}+B_{a}k_{y}^{2} & 0\\
0 & A_{a}k_{x}^{2}+B_{a}k_{y}^{2}
\end{array}\right]  \, ,
\end{equation}
which appears only for $a=\{c2,c1,v1,c2v1\}$ and the parameters $A_a$ and $B_a$ 
are the effective mass parameters obtained from the second-order perturbation theory. 
In the Supplemental Material\cite{supmat} we provide the specific definitions of these 
parameters in Sec.~II and in Sec.~III we write the three different $k \cdot p$ models explicitly.

\begin{table}[h!]
\caption{Values of the $k \cdot p$ parameters for the different $k \cdot p$ models. The parameters 
$P$ and $\alpha$ have units of eV$\cdot\textrm{\AA}$ and the parameters $A$ and $B$ 
are dimensionless.}
\begin{center}
{\renewcommand{\arraystretch}{1.2}
\begin{tabular*}{1.0\columnwidth}{@{\extracolsep{\fill}}
cccc}
\hline
\hline
 & Nph4 & ph4 & ph6\tabularnewline
\hline
$A_{c1}$        &  0.8696 &  0.8208 &  0.8405 \tabularnewline
$B_{c1}$        &  4.1667 &  0.5453 &  0.4800 \tabularnewline
$A_{v1}$        & -0.1372 & -0.1614 & -0.0958 \tabularnewline
$B_{v1}$        & -4.1667 & -0.4588 & -0.2926 \tabularnewline
$P_{c1v1}$      &         &  5.3696 &  5.4473 \tabularnewline
$\alpha_{c1v1}$ &         & -0.0195 & -0.0218 \tabularnewline
$P_{c2c1}$      &         &         &  0.2051 \tabularnewline
$\alpha_{c2c1}$ &         &         &  0.0070 \tabularnewline
$A_{c2}$        &         &         &  7.0622 \tabularnewline
$B_{c2}$        &         &         &  0.4845 \tabularnewline
$A_{c2v1}$      &         &         &  1.9669 \tabularnewline
$B_{c2v1}$      &         &         &  0.8598 \tabularnewline
\hline
\hline
\end{tabular*}}
\end{center}
\label{tab:kp_params}
\end{table}

To obtain the values for the different parameters that appear in the $k \cdot p$ Hamiltonians 
we perform a numerical fitting to the {\it ab initio} data. For a reliable description, 
we considered the band structure and the effective masses (as function of the wavevector $k$) 
along multiple directions of the first Brillouin zone starting 
from the $\Gamma$-point, namely $\Gamma$-X (along $k_x$), $\Gamma$-Y (along $k_y$) 
and $\Gamma$-S (which combines $k_x$ and $k_y$). These different constraints (band 
structure, effective masses and multiple directions) are used simultaneously in 
the fitting procedure, implemented via the LMFIT package in Python\cite{lmfit,FariaJunior2016PRB}. 
The use of the $k$-dependent effective mass calculated from {\it ab initio} weights 
the points in the vicinity of the $\Gamma$-point. Furthermore, special care must 
be taken for the linear-in-k terms, specially the alpha parameters. From {\it ab initio} 
we calculate the dipole matrix elements without SOC (within the optic code in WIEN2k\cite{Draxl2006CPC}), 
found $P_{c1v1}$ = 5.3230 eV$\cdot\textrm{\AA}$ and $P_{c2c1}$ = 0.1757 eV$\cdot\textrm{\AA}$. 
For the $\alpha$ parameter, we can estimate its value by comparing the change in 
the effective mass with and without SOC within the ab initio calculations. This 
approach gives as a range of values of $\alpha_{c1v1} \sim 0.007 - 0.015$ eV$\cdot\textrm{\AA}$. 
A detailed description of the evaluation of $\alpha_{c1v1}$ can be found in in Sec.~IV 
of the Supplemental Material\cite{supmat}. We emphasize that for each of the three different 
$k \cdot p$ models (ph6, ph4 and Nph4) we performed an individual fitting of the band 
structure and effective masses with double precision resulting in the parameter 
sets shown in Table~\ref{tab:kp_params}. From the Nph4 parameters we can readily 
obtain the effective masses: $m_{c1,x} = 1/A_{c1} = 1.15$, $m_{c1,y} = 1/B_{c1} = 0.24$, 
$ m_{v1,x} = 1/A_{v1} = 7.29$ and $m_{v1,y} = 1/B_{v1} = 0.24$. For the fitted 
values of $P_{c1v1}$ and $P_{c2c1}$ we our values shown in Table~\ref{tab:kp_params} 
are relatively close to the calculated by ab initio, thus showing that our fitting 
scheme is quite reliable. We point out that this small discrepancy between the fitted 
and calculated values from {\it ab initio} is well known within the $k \cdot p$ framework 
that the parameters are slightly modified to accommodate the band structure features. 
We note that the signs of the $\alpha$ parameters given in Table~\ref{tab:kp_params} 
can be reversed without affecting the band structure.

\begin{figure}[h!]
\begin{center} 
\includegraphics{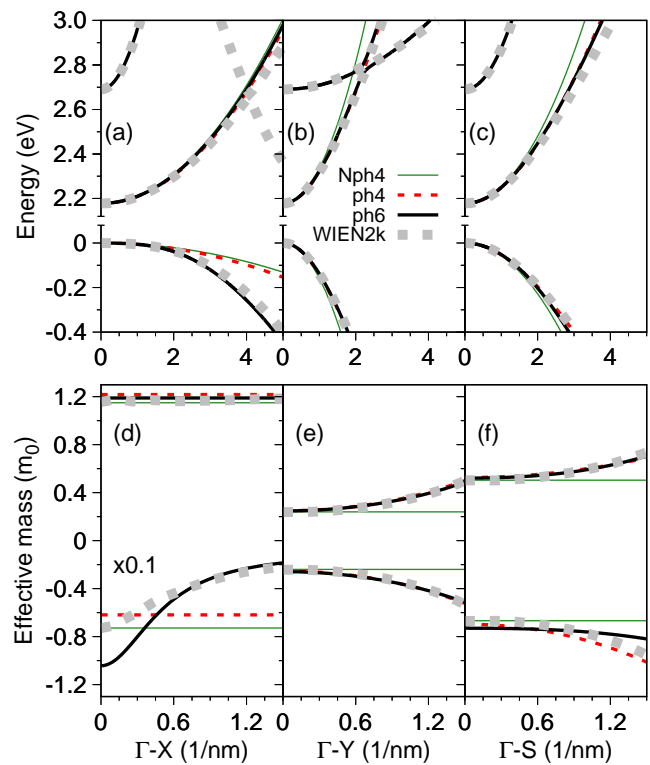}
\caption{(Color online) Calculated (a-c) band structures and (d-f) effective masses 
(shown for $c1$ and $v1$ only) for the different $k \cdot p$ models compared to the {\it ab initio} 
WIEN2k data along different directions in the first Brillouin zone starting from 
the $\Gamma$-point. In Fig.~\ref{fig:kp_bs_em}(d), the effective masses of valence 
band v1 are multiplied by a factor of 0.1. For the band structure (effective mass) 
x-axis, the reciprocal space distance of 5/nm (1.5/nm) corresponds to a percetage 
of $\sim 52 (16)$ for $\Gamma$-X, $\sim 73 (22)$ for $\Gamma$-Y and $\sim 43 (13)$ 
for $\Gamma$-S.}
\label{fig:kp_bs_em}
\end{center}
\end{figure}

The comparison between the fitted $k \cdot p$ models and the {\it ab initio} data 
is shown in Fig.~\ref{fig:kp_bs_em}. Up to 1.5 $\text{nm}^{-1}$, the limit in $k$-space 
used in our fitting, we find a good agreement for the band structure of all $k \cdot p$ 
models considered. For the effective mass analysis, we focus on $v1$ and $c1$ bands. 
We notice that Nph4 shows a constant dependence in all directions but agrees well 
very close to the $\Gamma$-point (as a parabolic dispersion should), ph4 also shows 
a nearly-constant dependence along $\Gamma$-X but now acquires a a dependence along 
 $\Gamma$-Y and $\Gamma$-S directions due to the interband couplings and, finally, 
ph6 provides a more realistic description of the effective mass k-dependence, with 
additional contributions due to the quadratic coupling between to $c2$-$c1$ bands 
and the interband coupling terms. Furthermore, although we enforced the $k$-limit 
of 1.5 $\text{nm}^{-1}$ for the fitting, we also find a good agreement with {\it ab initio} 
at even larger $k$-values for the band structure, except for v1 band along $\Gamma$-X 
in ph4 and Nph4 models. Within the ph6 model it is even possible to model the anticrossing 
of the $c2$ and $c1$ conduction bands along $\Gamma$-Y.

Let us now turn to additional properties that can be derived from the $k \cdot p$ models. 
As a consequence of including the interband SOC term (in ph4 and ph6 models), it 
is possible to compute the dipole strength between $v1$ and $c1$ bands not only for 
the armchair but also for the zigzag direction. The dipole strength between 
the top-most valence and bottom-most conduction bands as function of the wavevector 
$\vec{k}$ can be written as
\begin{equation}
\text{D}_{x(y)}(\vec{k})=\sum_{c,v}\left|\left\langle v,\vec{k}\left|\Pi_{x(y)}\right|c,\vec{k}\right\rangle \right|^{2} \, ,
\label{eq:Dxy}
\end{equation}
in which $x(y)$ refers to the zigzag (armchair) direction, the summation for the 
indices $c(v)$ takes into account both spin components of $c1(v1)$ bands and the 
$\vec{\Pi}$ operator is given in Eq.~(\ref{eq:Pi}). For the angular dependence of 
the dipole strength as a function of the transition energy, we can define the 
following quantity
\begin{align}
\text{D}(\theta,E) & =\sum_{c,v,\vec{k}}\left|\left\langle v,\vec{k}\left|\Pi_{y}\cos\theta+\Pi_{x}\sin\theta\right|c,\vec{k}\right\rangle \right|^{2}\nonumber \\
 & \qquad\quad\times\delta\left\{ E-\left[E_{c}(\vec{k})-E_{v}(\vec{k})\right]\right\} \, , 
\label{eq:Dtheta}
\end{align}
where $\theta$ is the angle defined as zero with respect to the armchair axis (y direction) 
and the $\vec{k}$ dependence in the summation take into account all possible transitions 
at a given transition energy of $E_c(\vec{k})-E_v(\vec{k})$. To compute such dipole 
strengths we use the values of $P$ and $\alpha$ given in Table~\ref{tab:kp_params}.

\begin{figure}[h!]
\begin{center} 
\includegraphics{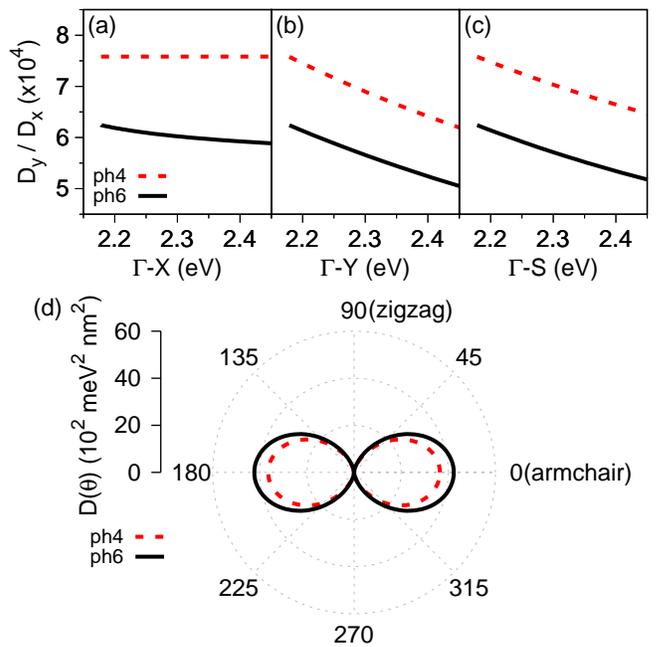}
\caption{(Color online) Ratio between armchair, D$_\text{y}$, and zigzag, D$_\text{x}$, 
dipole strength for ph4 and ph6 $k \cdot p$ models along (a) $\Gamma$-X, (b) $\Gamma$-Y 
and (c) $\Gamma$-S directions. In the x-axis we show the energy dependence by mapping 
the wavevector $\vec{k}$ to the transition energy $E_c(\vec{k})-E_v(\vec{k})$. 
(d) Angular dependence of the dipole strength for a transition energy of 5 meV above 
the band gap, calculated via Eq.~(\ref{eq:Dtheta}).}
\label{fig:kp_dipole}
\end{center}
\end{figure}

We discuss in Fig.~\ref{fig:kp_dipole} the dipole strength features of the ph4 
and ph6 $k \cdot p$ Hamiltonians. In Figs.~\ref{fig:kp_dipole}(a-c) we show the calculated 
dipole strength ratio between armchair and zigzag directions [$\text{D}_y$ / $\text{D}_x$, 
obtained from Eq.~(\ref{eq:Dxy})], as function of the transition energy, mapping 
the wavevector $\vec{k}$ to the transition energy $E_c(\vec{k})-E_v(\vec{k})$, 
along different directions of the Brillouin zone. We found that this ratio is quite 
large (4 to 5 orders of magnitude) but it is nonetheless nonzero, a feature that 
can only be achieved with the inclusion of the interband SOC terms (the $\alpha$ 
parameters). Moreover, this ratio for ph4 and ph6 are slighly different but both 
models (with their respective parameter sets) provide nearly the same trends. 
A complementary, and perhaps more instructive, way to visualize the dipole strength 
is to look at its angular dependence for a fixed transition energy using Eq.~(\ref{eq:Dtheta}). 
For a transition energy of 5 meV above the band gap, we show in Fig.~\ref{fig:kp_dipole}(d) 
the angular behavior of the dipole strength assuming the armchair direction at $\theta=0$. 
The 2-fold behavior of this angular dependence clearly reflects the symmetry 
of the phosphorene lattice. From the experimental perspective, recent photoluminescence 
measurements by Wang et al.\cite{Wang2015NatNano} reports that the emission along 
zigzag is consistently less than 3\% of the emission along armchair while in the 
study by Xu et al.\cite{Xu2016AdvMat}, the photoluminescence intensity as function 
of the polarization angle [Fig.~3(c)] suggests that the ratio between armchair and 
zigzag directions is two orders of magnitude or more. Based on our findings of $\text{D}_y$ / $\text{D}_x$ 
which take into account the intrinsic selection rules in phosphorene, the reduced 
dipole ratio observed in experiment might be associated to more complex phenomena 
of many-body interactions and carrier relaxation.

\begin{figure}[h!]
\begin{center} 
\includegraphics{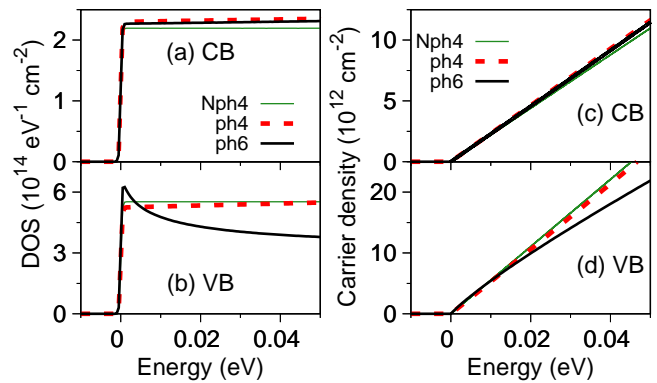}
\caption{(Color online) Calculated density of states for the different $k \cdot p$ models 
for (a) conduction and (b) valence bands. Comparison between the carrier density 
as function of energy for the different $k \cdot p$ models for (c) conduction and 
(d) valence bands. In the subfigures (a) and (c), the energy range starts from the band gap.}
\label{fig:kp_DOS_n}
\end{center}
\end{figure}

It is also interesting to investigate the behavior of the effective $k \cdot p$ models 
by calculating the density of states (DOS) and carrier density, physical quantities 
that play an important role in transport experiments. We show in Figs.~\ref{fig:kp_DOS_n}(a-b) 
the DOS for conduction and valence bands, respectively. As we noted previously for 
the band structure and effective masses shown in Fig.~\ref{fig:kp_bs_em}, the main 
differences among the $k \cdot p$ models appear in the valence band. Specifically, 
Nph4 provides a constant DOS dispersion, ph4 a nearly-constant behavior and ph6 
shows a higher value close to the band edge which decreases slightly as the energy 
increases, also observed in the {\it ab initio} calculations in Fig.1(a) of Ref.~[\onlinecite{Rudenko2015PRB}]. 
This is a typical feature observed in systems with reduced dimensionality and large 
effective mass, for instance in the valence band of conventional quantum wells based 
on zinc-blende GaAs\cite{FariaJunior2015PRB} and wurtzite GaN\cite{Chuang1997SST}.
By integrating the DOS we can obtain the carrier density as function of the 
energy, and similar trends of the DOS can also be seen in Figs.~\ref{fig:kp_DOS_n}(c-d) 
for conduction and valence band, respectively.


\section{Effective \MakeLowercase{g}-factors and Landau levels}
\label{sec:bmag}

In this section we incorporate the influence of external magnetic fields within 
the $k \cdot p$ models to investigate the effective g-factors and the LL spectra. 
Let us first start with the calculation of the effective g-factors, following the 
conventional perturbative approach within the $k \cdot p$ framework\cite{Roth1959PR,Hermann1977PRB,LewYanVoon2009}. 
We focus on magnetic fields that are oriented out of the monolayer plane, i. e., 
along $z$ direction, following the coordinate system shown in the inset of Fig.~\ref{fig:abinitio}. 
Under these conditions, the effective g-factor can be generally written as
\begin{equation}
g_n = g_0-i\frac{2m_{0}}{\hbar^{2}}\underset{l\neq n}{\sum}\frac{\Pi_{x}^{nl}\Pi_{y}^{ln}-\Pi_{y}^{nl}\Pi_{x}^{ln}}{E_{n}-E_{l}}
\label{eq:gfactor}
\end{equation}
in which $g_0$ is the bare electron g-factor, $n$ is the band of interest, $l$ runs 
over the other bands in the $k \cdot p$ model, the energy values in the denominator are 
the values at $\Gamma$-point, the $\vec{\Pi}$ operator is defined in Eq.~\ref{eq:Pi} and 
$\Pi^{nl}_{x(y)} = \left\langle n \left|\Pi_{x(y)}\right| l \right\rangle$, which 
follows the same form of the interband dipole coupling shown in Eq.~(\ref{eq:Dxy}).

\begin{table}[h!]
\caption{Calculated values for the effective g-factors using the ph4 and ph6 $k \cdot p$ models. 
The values in parentheses indicate the calculated g-factors with 
reversed signs of the $\alpha$ parameters given in Table~\ref{tab:kp_params}.}
\begin{center}
{\renewcommand{\arraystretch}{1.2}
\begin{tabular*}{0.8\columnwidth}{@{\extracolsep{\fill}}
ccc}
\hline 
\hline
 & ph4 & ph6\tabularnewline
\hline 
$g_{v1}$ & 2.0276 (1.9770) & 2.0309 (1.9737) \tabularnewline
$g_{c1}$ & 2.0276 (1.9770) & 2.0295 (1.9752) \tabularnewline
$g_{c2}$ &                 & 2.0009 (2.0038) \tabularnewline
\hline
\hline
\end{tabular*}}
\end{center}
\label{tab:gfactors}
\end{table}

Evaluating Eq.~(\ref{eq:gfactor}) specifically for the ph4 $k \cdot p$ model, the top-most 
valence and bottom-most conduction band g-factors read
\begin{equation}
g_{v1}=g_{c1}=g_{0}-2\left(\frac{2m_{0}}{\hbar^{2}}\right)\left(\frac{P_{c1v1}\alpha_{c1v1}}{E_{g}}\right) \, ,
\label{eq:gph4}
\end{equation}
while for the ph6 $k \cdot p$ model, the g-factors are given by
\begin{equation}
g_{v1}=g_{0}-2\left(\frac{2m_{0}}{\hbar^{2}}\right)\left(\frac{P_{c1v1}\alpha_{c1v1}}{E_{g}}\right) \, ,
\label{eq:gv1ph6}
\end{equation}
\begin{equation}
g_{c1}=g_{0}-2\left(\frac{2m_{0}}{\hbar^{2}}\right)\left(\frac{P_{c1v1}\alpha_{c1v1}}{E_{g}}+\frac{P_{c2c1}\alpha_{c2c1}}{E_{c2}-E_{g}}\right)
\label{eq:gc1ph6}
\end{equation}
and
\begin{equation}
g_{c2}=g_{0}-2\left(\frac{2m_{0}}{\hbar^{2}}\right)\left(\frac{P_{c2c1}\alpha_{c2c1}}{E_{c2}-E_{g}}\right)  \, ,
\label{eq:gc2ph6}
\end{equation}
with the parameters $P$ and $\alpha$ (discussed in Sec.~\ref{sec:kp}) with values 
given in Table~\ref{tab:kp_params} were used to compute the g-factor values. We 
emphasize that without including the interband SOC term (given by the $\alpha$ parameters) 
there is no correction to the effective g-factors from the bare electron g-factor, $g_0$.

Evaluating the g-factors given in Eqs.~(\ref{eq:gph4}), (\ref{eq:gv1ph6}), (\ref{eq:gc1ph6}) and (\ref{eq:gc2ph6}) 
using the parameters given in Table~\ref{tab:kp_params}, we show in Table~\ref{tab:gfactors} 
our predicted values for the effective g-factors within the ph4 and ph6 models. 
Due to the small value of $\alpha$, the corrections to the bare electron g-factor 
are of the order of $10^{-2}$. Furthermore, the values obtained for $g_{v1}$ and $g_{c1}$ 
from the different $k \cdot p$ models are consistent with each other. For the second 
conduction band, $g_{c2}$ can be only be accounted within the ph6 model and 
our predicted value is nearly the bare electron g-factor. The available 
g-factors experimentally determined for few-layer phosphorene via transport experiments 
for the top-most valence band are $g_{v1} = $ 1.8 - 2.7 by Gillgren et al.\cite{Gillgren20152DMat} 
and $g_{v1} = 2.0 \pm 0.1$ by Li et al.\cite{Li2016NatNano}. From a theoretical perspective, 
the value of $g_{v1} = 2.14$ for monolayer black phosphorus has been determined 
by Zhou et al.\cite{Zhou2017PRB_gph}, however, the interband SOC parameter $\alpha_{c1v1}$ 
was not theoretically obtained but estimated from the experimental data of Ref.~[\onlinecite{Long20162DMat}], 
leading to a value of $\alpha_{c1v1} \sim 0.45 \; \textrm{eV}\cdot\textrm{\AA}$. 
For the conduction band, recent experiments by Yang et al.\cite{Yang2018NL} in few-layer 
black phosphorus found surprisingly large g-factors of 5.7$\pm$0.7 for filling factor 
$\nu = 1$ due to electron-electron interaction. The g-factor value drops to 2.5$\pm$0.1 
for filling factor $\nu = 7$ once screening effects take place. Given the uncertainty in the 
experimentally determined effective g-factors obtained in few-layer phosphorene 
samples, certainly with the influence of many-body interactions, our predicted values 
have the advantage that they were obtained from a full theoretical approach taking into 
account the intrinsic contributions of bare phosphorene monolayers.

\begin{figure}[h!]
\begin{center} 
\includegraphics{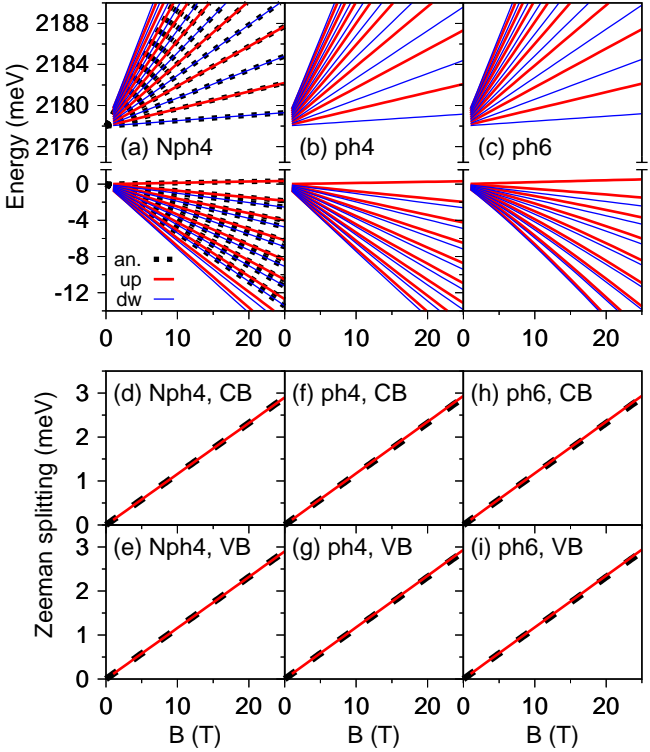}
\caption{(Color online) Calculated Landau levels for conduction and valence bands 
using the $k \cdot p$ model (a) Nph4, (b) ph4 and (c) ph6. Thick (thin) lines indicate 
spin up (down) LL branches. For the Nph4 calculations we also plot the analytical 
results of Eq.~(\ref{eq:LL_analytical}) in dashed lines. Zeeman splitting for the first 
Landau level in conduction and valence band for (d-e) Nph4, (f-g) ph4 and (h-i) ph6. 
Higher Landau levels follow the same behavior. Notice that without the interband 
interaction, the Zeeman splitting is the same as the bare electron, i. e., $\mu_B g_0 B$.}
\label{fig:LLs_ZS}
\end{center}
\end{figure}

Now let us turn to the LL spectra of the system, going beyond the effective 
g-factor approach by considering the envelope function approximation, 
combined with the minimal coupling and the Zeeman term\cite{LewYanVoon2009,Pryor2006PRL,vanBree2012PRB,FariaJunior2019PRB}. 
First, we notice that because of the lack of interband coupling in the Nph4 model, 
it is possible to find analytical solutions for the LLs\cite{Zhou2015SREP}, 
given by
\begin{equation}
E^{\pm}_{c1(v1)}(n)=E_{c1(v1)}+\left[M_{c1(v1)}\left(n+\frac{1}{2}\right)\pm\frac{g_{0}}{2}\right]\mu_{B}B \; ,
\label{eq:LL_analytical}
\end{equation}
with the subindex $c1(v1)$ denoting the bottom-most conduction (top-most valence) band, 
the superindex $\pm$ indicating the positive and negative Zeeman split LLs, 
$n=0,1,2,\ldots$ indicating the LL index, $E_{c1}=E_{g}$, $M_{c1}=2\sqrt{A_{c1}B_{c1}}$, $E_{v1}=0$ 
and $M_{v1}=-2\sqrt{A_{v1}B_{v1}}$. For the ph4 and ph6 models, we employ the numerical 
technique of the finite differences\cite{Chuang1997SST} to obtain the LL spectra. 
We also apply this numerical approach to the Nph4 model for comparison. We assumed 
$B=B\hat{z}$ and the vector potential given by $\vec{A}=Bx\hat{y}$. For the numerical 
discretization we considered the a system size of $L=200 \; \text{nm}$ with 401 points and 
hard-wall boundary conditions.

In Fig.~\ref{fig:LLs_ZS} we summarize the LL spectra and the ZS for 
the different $k \cdot p$ models with parameters from Table~\ref{tab:kp_params}. 
Let us start with the Nph4 model, with LLs shown in 
Fig.~\ref{fig:LLs_ZS}(a) and ZSs in Figs.~\ref{fig:LLs_ZS}(d-e). We found a very good 
agreement between our numerical calculations and the analytical approach of Eq.~(\ref{eq:LL_analytical}), 
however, due to the lack of interband coupling, the ZS is given only by the bare 
electron g-factor. For ph4 and ph6 models LL spectra, shown in Figs.~\ref{fig:LLs_ZS}(b-c) 
respectively, the situation is quite similar with just different energy separations 
between LLs arising due to the different coupling in the Hamiltonians. Turning to 
the ZS, shown in Figs.~\ref{fig:LLs_ZS}(f-i), we observe that both ph4 and ph6 models 
slightly deviate from the bare electron g-factor case, as expected and already noticed in the 
effective g-factor case. These different features in the LLs and ZS available in the ph4 and ph6 models could 
be investigated as signatures in magneto-transport spectra\cite{Zhou2015SREP,Tahir2015PRB}.


\section{Excitons}
\label{sec:excitons}

\begin{figure}[h!]
\begin{center} 
\includegraphics{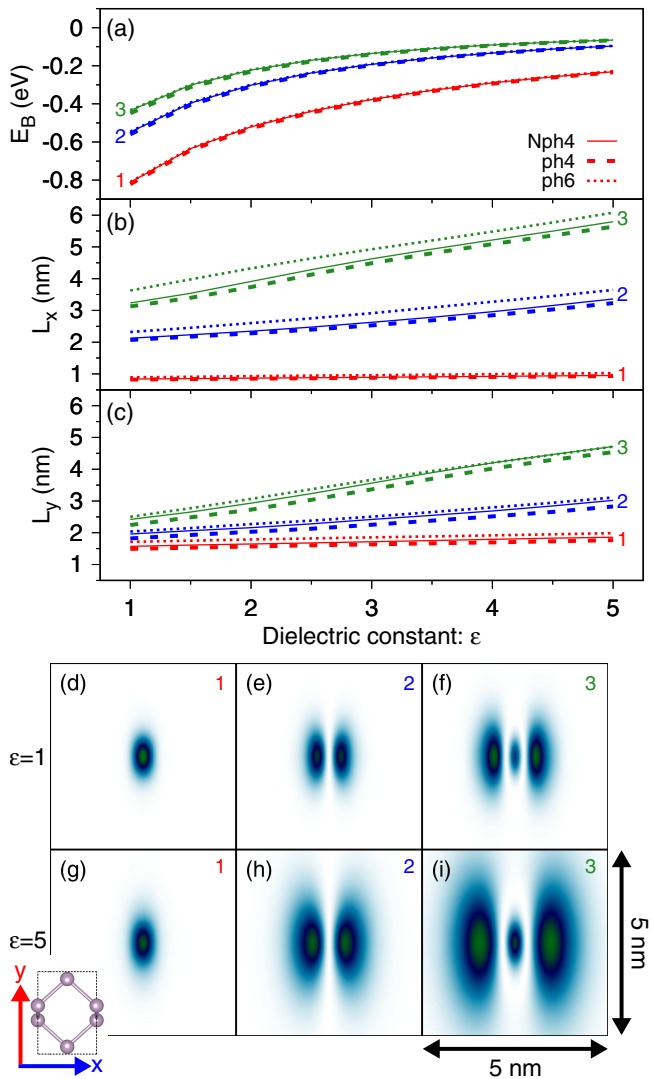}
\caption{(Color online) Calculated exciton (a) biding energy, $\text{E}_\text{B}$, 
(b) diameter along zigzag direction, $\text{L}_x$, and (c) diameter along armchair 
direction, $\text{L}_y$, as function of the effective dielectric constant 
for the lowest three excitonic states using the different $k \cdot p$ models. The zero energy 
in subfigure (a) is the single-particle band gap of phosphorene of 2.178 eV. The 
exciton probability densities in real space are presented for (d-f) $\varepsilon=1$ and 
(g-i) $\varepsilon=5$ using the Nph4 model. The inset in the bottom left corner 
shows the top view of phosphorene unit cell (not in scale).}
\label{fig:excitons}
\end{center}
\end{figure}

Since phosphorene is a direct band gap semiconductor with interesting optical properties, 
it is also important to check the consistency of our effective $k \cdot p$ models 
and parameters by calculating the excitonic spectra. Using the effective BSE\cite{Rohlfing2000PRB,Scharf2016PRB,Tedeschi2018PRBr}, 
we focus on direct excitons at zero temperature assuming the electron-hole interaction 
to be mediated by the Rytova-Keldysh potential\cite{Rytova1967, Keldysh1979, Cudazzo2011PRB}, 
given by 
\begin{equation}
\text{v}(r)=\frac{e^{2}}{8\varepsilon_{0}r_{0}}\left[H_{0}\left(\frac{\varepsilon}{r_{0}}r\right)-Y_{0}\left(\frac{\varepsilon}{r_{0}}r\right)\right] \, ,
\end{equation}
in which $H_{0}$ is the zeroth-order Struve function, $Y_{0}$ is the zeroth-order 
Bessel function of the second kind, $r=\sqrt{x^{2}+y^{2}}$, $e$ is the electron 
charge, $\varepsilon_0$ is the vacuum permittivity, $\varepsilon=(\varepsilon_t+\varepsilon_b)/2$ 
is the effective dielectric constant given by the average of top, $\varepsilon_t$, 
and bottom, $\varepsilon_b$, dielectric constants, and $r_{0}=2\pi\zeta$ with $\zeta$ 
being the 2D dielectric susceptibility with typical values in the literature of 
$\zeta=4.1\;\textrm{\AA}$\cite{Rodin2014PRB}, $\zeta=3.85\;\textrm{\AA}$\cite{Seixas2015PRB} and 
$\zeta=3.8\;\textrm{\AA}$\cite{Prada2015PRB}. In our calculations we assume 
$\zeta=4\;\textrm{\AA}$, leading to a value of $r_{0}\approx25\;\textrm{\AA}$.

We analyze the effect of the dielectric environment considering different values 
of $\varepsilon$ from 1 to 5, in order to cover a reasonable range of experimental 
realizations of phosphorene. For instance, we identify three important cases that 
fall within the range of $\varepsilon$ values we considered: 
(i) freestanding phosphorene is equivalent to $\varepsilon=1$ ($\varepsilon_t=\varepsilon_b=1$); 
(ii) phosphorene on SiO$_\text{2}$ substrate is equivalent to $\varepsilon=2.45$ 
($\varepsilon_t=1$ and $\varepsilon_b=3.9$\cite{Berkelbach2013PRB}); and (iii) 
encapsulated phosphorene with boron nitride is equivalent to $\varepsilon=4.5$ 
($\varepsilon_t=\varepsilon_b=4.5$\cite{Stier2018PRL}). For all the $k \cdot p$ models we 
solve the BSE numerically using a 2D $k$-grid of -0.5 to 0.5 $\textrm{\AA}^{-1}$ 
in $k_x$ (zigzag) and -0.3 to 0.3 $\textrm{\AA}^{-1}$ in $k_y$ (armchair) with a 
total discretization of $(2\text{N}_x+1)\times(2\text{N}_y+1)$ with $(\text{N}_x,\text{N}_y)=\left[(70,42),(60,36)\right]$
The relative error between these two mesh discretizations is $\sim 1\%$. The final 
values of exciton biding energies and diameters are then obtained using a linear extrapolation 
of the values calculated in these two grid sizes. We define the exciton diameter 
as the full width at half maximum of the exciton probability density in real space.

In Fig.~\ref{fig:excitons} we summarize our findings for the excitonic spectra focusing 
on the lowest 3 excitonic states. We show in Figs.~\ref{fig:excitons}(a)-(c) the 
exciton biding energy, $\text{E}_\text{B}$ (measured from the single-particle band 
gap), the exciton diameter along the zigzag direction, $\text{L}_x$, and the exciton 
diameter along the armchair direction, $\text{L}_y$, as function of the effective 
dielectric constant, respectively. As an initial remark, all the models are in 
reasonable quantitative agreement with each other and thus it suffices to discuss 
the general behavior of $\text{E}_\text{B}$, $\text{L}_x$ and $\text{L}_y$ as function 
of $\varepsilon$ irrespective of the model. Increasing the value of $\varepsilon$, 
we show that the magnitude of $\text{E}_\text{B}$ decreases in a nonlinear fashion whereas 
$\text{L}_x$ and $\text{L}_y$ increase linearly, with increasing slope for higher 
exciton levels. This behavior is in qualitative agreement with 2D excitons obtained 
from the Wannier equation and the typical Coulomb potential ($\text{E}_\text{B} \sim \varepsilon^{-2}$ 
and $\text{R} \sim \varepsilon$)\cite{Haug2004}. In Figs.~\ref{fig:excitons}(d)-(f) and 
Figs.~\ref{fig:excitons}(g)-(i) we show the probability density of the exciton wavefunction 
in real space for $\varepsilon=1$ and $\varepsilon=5$, respectively. We can identify 
the first exciton to a 1s state, the second exciton to a 2p$_\text{x}$ and the third exciton 
to a 2s state. This exciton ordering is in agreement with recent {\it ab initio} calculations\cite{Qiu2017NL}.
We emphasize that although Nph4 provides a consistent description of excitonic 
effects, it does not allow any further calculations involving dipole coupling 
[see Fig.~\ref{fig:kp_dipole}] and therefore one should rely on either 
the ph4 or the ph6 models. From the exciton perspective we revisit the dipole ratio 
between armchair and zigzag directions for the first exciton level and found the 
values of $6.4 \times 10^{4}$ for ph4 and $5.1 \times 10^{4}$ for ph6, thus showing 
the same order of magnitude as the values calculated in Fig.~\ref{fig:kp_dipole}.



Finally, let us compare our calculated exciton biding energies with the literature. 
For a freestanding mononalyer phosphorene we find $\text{E}_\text{B}$ $\sim$0.81 eV, 
in good agreement with the range found in the literature 0.75 - 0.86 eV \cite{Chaves2016PRB, Prada2015PRB, Rodin2014PRB, Tran2014PRB, Tran20152DMat, Hunt2018PRB, Frank2019PRX}.
For a monolayer phosphorene on SiO$_\text{2}$ substrate we obtained $\text{E}_\text{B}$ $\sim$0.44 eV, 
consistent with the reported values of 0.38 eV\cite{CastellanosGomez20142DMat}, 
0.4 eV\cite{Rodin2014PRB} and 0.46 eV\cite{Hunt2018PRB}. And for encapsulated 
phosphorene with boron nitride, we found $\text{E}_\text{B}$ $\sim$0.26 eV, also 
in good agreement with reported values of $\sim$0.22 eV\cite{Rodin2014PRB} and 
$\sim$0.26 eV\cite{Hunt2018PRB}. 


\section{Conclusions}
\label{sec:conclusions}

In summary, we have developed effective four- and six-band $k \cdot p$ Hamiltonians exploiting the full 
symmetry of monolayer phosphorene by including the interband spin-orbit coupling interaction, 
a term previously neglected in the literature and of crucial importance for the 
treatment of anisotropic 2D materials. Specifically, the inclusion of such spin-orbit 
term allows the calculation of two important features in phosphorene: (i) the proper description 
of the interband dipole interaction not only along armchair but also along zigzag 
directions and (ii) the estimation of the effective g-factors and the Zeeman splittings 
of Landau levels. To obtain the $k \cdot p$ parameters, we have performed a systematic fitting 
of the band structure and the $k$-dependence of the effective masses to reliable {\it ab initio} 
calculations. Our $k \cdot p$ models highlight the intrinsic characteristics of 
monolayer phosphorene and were investigated in light of different physical aspects 
showing a correct description of the dipole selection rules, effective g-factors, 
stable behavior of the Landau level spectra and consistent description of the excitonic 
spectra (in good agreement with reported values). Furthermore, our findings suggest 
that one must be careful in comparing the calculated intrinsic g-factors to the available experimental 
data since many-body effects could have an important contribution. Finally, the presented 
$k \cdot p$ Hamiltonians and parameters can be directly applied to investigate many-body 
effects, transport phenomena in the presence of magnetic field, optical properties 
including excitonic effects and can be possibly coupled to other available Hamiltonians 
of 2D materials to investigate novel van der Waals heterostructures. 

{\it Note added in proof.} After this manuscript was completed we became aware of 
Ref.~[\onlinecite{Li2019arXiv}] which also calculates the interband spin-orbit coupling 
parameter $\alpha$ using a different theoretical approach.


\section*{Acknowledgments}

The authors acknowledge financial support from the Alexander von Humboldt Foundation, 
Capes (grant No. 99999.000420/2016-06), DFG SFB 1177 (A09, B05), National Science 
Centre under the contract DEC-2018/29/B/ST3/01892 and VVGS-2019-1227. We acknowledge 
anonymous referees for valuable feedback to this manuscript. We thank Pengke Li 
for helpful discussions.



\clearpage

\onecolumngrid

\section*{Supplemental Material for the paper ``$k \cdot p$ theory for phosphorene: effective g-factors, Landau levels, and excitons''}

\vspace{1cm}


\section*{\uppercase{I.} Symmetry properties of phosphorene}

\begin{figure}[h!]
\begin{center} 
\includegraphics{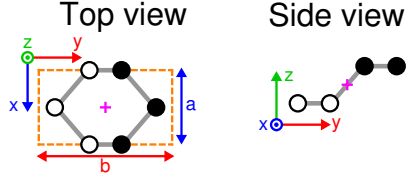}
\caption{(Color online) Top and side view of the unit cell in direct space. The 
closed (open) circles indicate the top (bottom) phosphorus atoms and the cross 
indicate the origin of the coordinate system.}
\label{fig:ph_ucell}
\end{center}
\end{figure}

In this section we briefly discuss the symmetry properties of phosphorene. The unit cell 
in direct space and the coordinate system is shown in Fig.~\ref{fig:ph_ucell}. 
In Table~\ref{tab:D2h} we present the character table for the nonsymmorphic group 
$D_{2h}$ of phosphorene. The symmetry operations are as follows: $E$ is the identity, 
$C_{a}$ is a rotation of $\pi/2$ along the $a=x,y,z$ axis, $I$ is the inverstion 
and $\tau$ is the translation vector given by $\vec{\tau} = \frac{a}{2}\hat{x} + \frac{b}{2}\hat{y}$. 
The multiplication table of the irreducible represenations are given in the Table~\ref{tab:D2h_mult}. 
Please refer to the study of Li and Appelbaum\cite{Li2014PRB} ...

\begin{table}[h!]
\caption{Character table for the nonsymmorphic group $D_{2h}$ of phosphorene.}
\begin{center}
{\renewcommand{\arraystretch}{1.2}
\begin{tabular*}{1.0\columnwidth}{@{\extracolsep{\fill}}
cccccccccc}
\hline
\hline
 & $E$ & $\left\{ C_{z}|\vec{\tau}\right\} $ & $C_{x}$ & $\left\{ C_{y}|\vec{\tau}\right\} $ & $I$ & $\left\{ IC_{z}|\vec{\tau}\right\} $ & $IC_{x}$ & $\left\{ IC_{y}|\vec{\tau}\right\} $ & functions\tabularnewline
\hline
$\Gamma_{1+}$ & 1 & 1 & 1 & 1 & 1 & 1 & 1 & 1 & $x^{2},y^{2},z^{2}$\tabularnewline
$\Gamma_{x+}$ & 1 & -1 & 1 & -1 & 1 & -1 & 1 & -1 & $R_{x},yz$\tabularnewline
$\Gamma_{z+}$ & 1 & 1 & -1 & -1 & 1 & 1 & -1 & -1 & $R_{z},xy$\tabularnewline
$\Gamma_{y+}$ & 1 & -1 & -1 & 1 & 1 & -1 & -1 & 1 & $R_{y},zx$\tabularnewline
\hline
$\Gamma_{1-}$ & 1 & 1 & 1 & 1 & -1 & -1 & -1 & -1 & $xyz$\tabularnewline
$\Gamma_{x-}$ & 1 & -1 & 1 & -1 & -1 & 1 & -1 & 1 & $x$\tabularnewline
$\Gamma_{z-}$ & 1 & 1 & -1 & -1 & -1 & -1 & 1 & 1 & $z$\tabularnewline
$\Gamma_{y-}$ & 1 & -1 & -1 & 1 & -1 & 1 & 1 & -1 & $y$\tabularnewline
\hline 
\hline
\end{tabular*}}
\end{center}
\label{tab:D2h}
\end{table}

\begin{table}[h!]
\caption{Multiplication table for the irreps of the character table given in Table \ref{tab:D2h}.}
\begin{center}
{\renewcommand{\arraystretch}{1.2}
\begin{tabular*}{1.0\columnwidth}{@{\extracolsep{\fill}}
ccccccccc}
\hline
\hline
 & $\Gamma_{1+}$ & $\Gamma_{x+}$ & $\Gamma_{z+}$ & $\Gamma_{y+}$ & $\Gamma_{1-}$ & $\Gamma_{x-}$ & $\Gamma_{z-}$ & $\Gamma_{y-}$\tabularnewline
\hline
$\Gamma_{1+}$ & $\Gamma_{1+}$ & $\Gamma_{x+}$ & $\Gamma_{z+}$ & $\Gamma_{y+}$ & $\Gamma_{1-}$ & $\Gamma_{x-}$ & $\Gamma_{z-}$ & $\Gamma_{y-}$\tabularnewline
$\Gamma_{x+}$ & $\Gamma_{x+}$ & $\Gamma_{1+}$ & $\Gamma_{y+}$ & $\Gamma_{z+}$ & $\Gamma_{x-}$ & $\Gamma_{1-}$ & $\Gamma_{y-}$ & $\Gamma_{z-}$\tabularnewline
$\Gamma_{z+}$ & $\Gamma_{z+}$ & $\Gamma_{y+}$ & $\Gamma_{1+}$ & $\Gamma_{x+}$ & $\Gamma_{z-}$ & $\Gamma_{y-}$ & $\Gamma_{1-}$ & $\Gamma_{x-}$\tabularnewline
$\Gamma_{y+}$ & $\Gamma_{y+}$ & $\Gamma_{z+}$ & $\Gamma_{x+}$ & $\Gamma_{1+}$ & $\Gamma_{y-}$ & $\Gamma_{z-}$ & $\Gamma_{x-}$ & $\Gamma_{1-}$\tabularnewline
\hline
$\Gamma_{1-}$ & $\Gamma_{1-}$ & $\Gamma_{x-}$ & $\Gamma_{z-}$ & $\Gamma_{y-}$ & $\Gamma_{1+}$ & $\Gamma_{x+}$ & $\Gamma_{z+}$ & $\Gamma_{y+}$\tabularnewline
$\Gamma_{x-}$ & $\Gamma_{x-}$ & $\Gamma_{1-}$ & $\Gamma_{y-}$ & $\Gamma_{z-}$ & $\Gamma_{x+}$ & $\Gamma_{1+}$ & $\Gamma_{y+}$ & $\Gamma_{z+}$\tabularnewline
$\Gamma_{z-}$ & $\Gamma_{z-}$ & $\Gamma_{y-}$ & $\Gamma_{1-}$ & $\Gamma_{x-}$ & $\Gamma_{z+}$ & $\Gamma_{y+}$ & $\Gamma_{1+}$ & $\Gamma_{x+}$\tabularnewline
$\Gamma_{y-}$ & $\Gamma_{y-}$ & $\Gamma_{z-}$ & $\Gamma_{x-}$ & $\Gamma_{1-}$ & $\Gamma_{y+}$ & $\Gamma_{z+}$ & $\Gamma_{x+}$ & $\Gamma_{1+}$\tabularnewline
\hline
\hline 
\end{tabular*}}
\end{center}
\label{tab:D2h_mult}
\end{table}


\section*{\uppercase{II.} Definition of k.p parameters}

In this section we present the definitions of the k.p parameters that appear in 
the Hamiltonians in the main text. The unperturbed terms are given by
\begin{equation}
E_{c2}=\left\langle \Gamma_{x+}^{c2}\left|\mathbf{H_{0}}\right|\Gamma_{x+}^{c2}\right\rangle \, ,
\end{equation}
\begin{equation}
E_{c1}=\left\langle \Gamma_{z-}^{c1}\left|\mathbf{H_{0}}\right|\Gamma_{z-}^{c1}\right\rangle  \, ,
\end{equation}
\begin{equation}
\left\langle \Gamma_{x+}^{v1}\left|\mathbf{H_{0}}\right|\Gamma_{x+}^{v1}\right\rangle = 0 \, ,
\end{equation}
in which $\mathbf{H_{0}}$ is the unperturbed term of the Hamiltonian.

For the terms that appear linear in k, we have
\begin{equation}
P_{c2c1}=i\frac{\hbar}{m_{0}}\left\langle \Gamma_{x+}^{c2}\left|p_{y}\right|\Gamma_{z-}^{c1}\right\rangle \, ,
\end{equation}
\begin{equation}
P_{c1v1}=i\frac{\hbar}{m_{0}}\left\langle \Gamma_{z-}^{c1}\left|p_{y}\right|\Gamma_{x+}^{v1}\right\rangle \, ,
\end{equation}
\begin{equation}
\alpha_{c2c1}=\frac{\hbar^{2}}{4m_{0}^{2}c^{2}}\left\langle \Gamma_{x+}^{c2}\left|\frac{\partial V_{C}}{\partial y}\right|\Gamma_{z-}^{c1}\right\rangle \, ,
\end{equation}
\begin{equation}
\alpha_{c1v1}=\frac{\hbar^{2}}{4m_{0}^{2}c^{2}}\left\langle \Gamma_{z-}^{c1}\left|\frac{\partial V_{C}}{\partial y}\right|\Gamma_{x+}^{v1}\right\rangle  \, ,
\end{equation}
in which $V_C$ is the crystal potential.

For the terms that appear with the quadratic k, we have
\begin{equation}
A_{c2}=1+\frac{2}{m_{0}}\sum_{\beta}^{B\left[\Gamma_{1-}\right]}\frac{\left|\left\langle \Gamma_{x+}^{c2}\left|p_{x}\right|\beta\right\rangle \right|^{2}}{E(\Gamma_{x+}^{c2})-E_{\beta}} \, ,
\end{equation}
\begin{equation}
B_{c2}=1+\frac{2}{m_{0}}\sum_{\beta}^{B\left[\Gamma_{z-}\right]}\frac{\left|\left\langle \Gamma_{x+}^{c2}\left|p_{y}\right|\beta\right\rangle \right|^{2}}{E(\Gamma_{x+}^{c2})-E_{\beta}} \, ,
\end{equation}
\begin{equation}
A_{c2v1}=\frac{2}{m_{0}}\sum_{\beta}^{B\left[\Gamma_{1-}\right]}\frac{\left\langle \Gamma_{x+}^{c2}\left|p_{x}\right|\beta\right\rangle \left\langle \beta\left|p_{x}\right|\Gamma_{x+}^{v1}\right\rangle }{E(\Gamma_{x+}^{c2},\Gamma_{x+}^{v1})-E_{\beta}} \, ,
\end{equation}
\begin{equation}
B_{c2v1}=\frac{2}{m_{0}}\sum_{\beta}^{B\left[\Gamma_{z-}\right]}\frac{\left\langle \Gamma_{x+}^{c2}\left|p_{y}\right|\beta\right\rangle \left\langle \beta\left|p_{y}\right|\Gamma_{x+}^{v1}\right\rangle }{E(\Gamma_{x+}^{c2},\Gamma_{x+}^{v1})-E_{\beta}} \, ,
\end{equation}
\begin{equation}
A_{c1}=1+\frac{2}{m_{0}}\sum_{\beta}^{B\left[\Gamma_{y+}\right]}\frac{\left|\left\langle \Gamma_{z-}^{c1}\left|p_{x}\right|\beta\right\rangle \right|^{2}}{E(\Gamma_{z-}^{c1})-E_{\beta}} \, ,
\end{equation}
\begin{equation}
B_{c1}=1+\frac{2}{m_{0}}\sum_{\beta}^{B\left[\Gamma_{x+}\right]}\frac{\left|\left\langle \Gamma_{z-}^{c1}\left|p_{y}\right|\beta\right\rangle \right|^{2}}{E(\Gamma_{z-}^{c1})-E_{\beta}} \, ,
\end{equation}
\begin{equation}
A_{v1}=1+\frac{2}{m_{0}}\sum_{\beta}^{B\left[\Gamma_{1-}\right]}\frac{\left|\left\langle \Gamma_{x+}^{v1}\left|p_{x}\right|\beta\right\rangle \right|^{2}}{E(\Gamma_{x+}^{v1})-E_{\beta}} \, ,
\end{equation}
\begin{equation}
B_{v1}=1+\frac{2}{m_{0}}\sum_{\beta}^{B\left[\Gamma_{z-}\right]}\frac{\left|\left\langle \Gamma_{x+}^{v1}\left|p_{y}\right|\beta\right\rangle \right|^{2}}{E(\Gamma_{x+}^{v1})-E_{\beta}} \, ,
\end{equation}
with the irreductible representations that contribute to nonzero matrix elements 
indicated in the square brackets above the summation.


\section*{\uppercase{III.} Hamiltonians}

Explicity matrix form of the Hamiltonian ph6:
\begin{equation}
\left\{ \left|\Gamma_{x+}^{c2}\uparrow\right\rangle ,\left|\Gamma_{x+}^{c2}\downarrow\right\rangle ,\left|\Gamma_{z-}^{c1}\uparrow\right\rangle ,\left|\Gamma_{z-}^{c1}\downarrow\right\rangle ,\left|\Gamma_{x+}^{v1}\uparrow\right\rangle ,\left|\Gamma_{x+}^{v1}\downarrow\right\rangle \right\} 
\end{equation}
{\small{}
\begin{equation}
\left[\begin{array}{cc|cc|cc}
E_{c2}+M_{c2} & 0 & -iP_{c2c1}k_{y}-\alpha_{c2c1}k_{x} & 0 & M_{c2v1} & 0\\
0 & E_{c2}+M_{c2} & 0 & -iP_{c2c1}k_{y}+\alpha_{c2c1}k_{x} & 0 & M_{c2v1}\\
\hline iP_{c2c1}k_{y}-\alpha_{c2c1}k_{x} & 0 & E_{c1}+M_{c1} & 0 & -iP_{c1v1}k_{y}-\alpha_{c1v1}k_{x} & 0\\
0 & iP_{c2c1}k_{y}+\alpha_{c2c1}k_{x} & 0 & E_{c1}+M_{c1} & 0 & -iP_{c1v1}k_{y}+\alpha_{c1v1}k_{x}\\
\hline M_{c2v1} & 0 & iP_{c1v1}k_{y}-\alpha_{c1v1}k_{x} & 0 & M_{v1} & 0\\
0 & M_{c2v1} & 0 & iP_{c1v1}k_{y}+\alpha_{c1v1}k_{x} & 0 & M_{v1}
\end{array}\right]
\end{equation}
}

Explicity matrix form of the Hamiltonian ph4:
\begin{equation}
\left\{ \left|\Gamma_{z-}^{c1}\uparrow\right\rangle ,\left|\Gamma_{z-}^{c1}\downarrow\right\rangle ,\left|\Gamma_{x+}^{v1}\uparrow\right\rangle ,\left|\Gamma_{x+}^{v1}\downarrow\right\rangle \right\} 
\end{equation}
\begin{equation}
\left[\begin{array}{cc|cc}
E_{c1}+M_{c1} & 0 & -iP_{c1v1}k_{y}-\alpha_{c1v1}k_{x} & 0\\
0 & E_{c1}+M_{c1} & 0 & -iP_{c1v1}k_{y}+\alpha_{c1v1}k_{x}\\
\hline iP_{c1v1}k_{y}-\alpha_{c1v1}k_{x} & 0 & M_{v1} & 0\\
0 & iP_{c1v1}k_{y}+\alpha_{c1v1}k_{x} & 0 & M_{v1}
\end{array}\right]
\end{equation}

Explicity matrix form of the Hamiltonian Nph4:
\begin{equation}
\left\{ \left|\Gamma_{z-}^{c1}\uparrow\right\rangle ,\left|\Gamma_{z-}^{c1}\downarrow\right\rangle ,\left|\Gamma_{x+}^{v1}\uparrow\right\rangle ,\left|\Gamma_{x+}^{v1}\downarrow\right\rangle \right\} 
\end{equation}
\begin{equation}
\left[\begin{array}{cc|cc}
E_{c1}+M_{c1} & 0 & 0 & 0\\
0 & E_{c1}+M_{c1} & 0 & 0\\
\hline 0 & 0 & M_{v1} & 0\\
0 & 0 & 0 & M_{v1}
\end{array}\right]
\end{equation}

To shorten the notation, the $M_{a}$ terms are defined as
\begin{equation}
M_{a}=\frac{\hbar^{2}}{2m_{0}}\left(A_{a}k_{x}^{2}+B_{a}k_{y}^{2}\right)
\end{equation}


\section*{\uppercase{IV.} Estimating the interband spin-orbit coupling}

The contribution to the effective mass in conduction
bands from $\alpha_{c1v1}$ in the ph4 model can be written as:

\begin{equation}
\frac{1}{m_{c1,x}^{*}}=\frac{\alpha_{c1v1}^{2}}{\left(\frac{\hbar^{2}}{2m_{0}}\right)E_{g}}
\end{equation}

The value $\frac{\alpha_{c1v1}^{2}}{\left(\frac{\hbar^{2}}{2m_{0}}\right)E_{g}}$
is of the order of the spin-orbit couping (SOC) correction to the
effective mass since no other parameters in the Hamiltonian would
contribute to that. Therefore, by knowing the effective mass with
and without SOC, we can estimate $\alpha_{c1v1}$ as:

\begin{equation}
\Delta M=\left|\left[\frac{1}{m^{*}}\right]_{\text{SOC}}-\left[\frac{1}{m^{*}}\right]_{\text{noSOC}}\right|
\end{equation}

\begin{equation}
\alpha_{c1v1}=\sqrt{\left(\frac{\hbar^{2}}{2m_{0}}\right)E_{g}\Delta M}
\end{equation}

Fitting of $1/m^{*}$ using the {\it ab initio}
calculations without and with SOC in double precision:

\begin{center}
\begin{tabular}{|l|c|c|c|c|c|}
\hline 
CB & 0.01/AA (1.05\%) & 0.02/AA (2.1\%) & 0.04/AA (4.2\%) & 0.06/AA (6.3\%) & 0.08/AA (8.4\%)\tabularnewline
\hline 
$\frac{1}{m_{c1,x}^{*}}$, noSOC & 0.65677678 & 0.76907818 & 0.79621609 & 0.79552087 & 0.79495243\tabularnewline
\hline 
$\frac{1}{m_{c1,x}^{*}}$, SOC & 0.65678272 & 0.76909900 & 0.79624126 & 0.79554921 & 0.79497681\tabularnewline
\hline 
$\Delta M$ & 0.00000594 & 0.00002082 & 0.00002517 & 0.00002834 & 0.00002439\tabularnewline
\hline 
$\alpha_{c1v1}$ (eV{*}AA) & 0.00701930 & 0.01314255 & 0.01445248 & 0.01533580 & 0.01422551\tabularnewline
\hline 
\end{tabular}
\end{center}



\begin{thebibliography}{99}
\bibitem{Liu2014ACSNano}
H. Liu, A. T. Neal, Z. Zhu, Z. Luo, X. Xu, D. Tománek, and P. D. Ye, 
ACS Nano {\bf 8}, 4033 (2014).

\bibitem{Lu2014NR}
W. Lu, H. Nan, J. Hong, Y. Chen, C. Zhu, Z. Liang, X. Ma, Z. Ni, C. Jin, and Z. Zhang, 
Nano Res. {\bf 7}, 853 (2014).

\bibitem{Liang2014NL}
L. Liang, J. Wang, W. Lin, B. G. Sumpter, V. Meunier, and M. Pan, 
Nano Lett. {\bf 14}, 6400 (2014).

\bibitem{Zhang2014ACSNano}
S. Zhang, J. Yang, R. Xu, F. Wang, W. Li, M. Ghufran, Y.-W. Zhang, Z. Yu, G. Zhang, Q. Qin, and Y. Lu,
ACS Nano {\bf 8}, 9590 (2014).

\bibitem{CastellanosGomez20142DMat}
A. Castellanos-Gomez, L. Vicarelli, E. Prada, J. O Island, K. L. Narasimha-Acharya, 
S. I Blanter, D. J. Groenendijk, M. Buscema, G. A. Steele, J. V. Alvarez, H. W. Zandbergen, 
J. J. Palacios, and H. S. J. van der Zan, 
2D Mater. {\bf 1}, 025001 (2014).

\bibitem{Qiao2014NatComm}
J. Qiao, X. Kong, Z.-X. Hu, F. Yang, and W. Ji, 
Nat. Commun. {\bf 5}, 4475 (2014).

\bibitem{Keyes1953}
R. W. Keyes, 
Phys. Rev. {\bf 92}, 580 (1953)

\bibitem{Das2014NL}
S. Das, W. Zhang, M. Demarteau, A. Hoffmann, M. Dubey, and A. Roelofs, 
Nano Lett. {\bf 14}, 5733 (2014).

\bibitem{Wang2015NatNano}
X. Wang, A. M. Jones, K. L. Seyler, V. Tran, Y. Jia, H. Zhao, H. Wang, L. Yang, X. Xu, and F. Xia, 
Nat. Nanotech. {\bf 10}, 517 (2015).

\bibitem{Frank2019PRX}
T. Frank, R. Derian, K. Tokar, L. Mitas, J. Fabian, and I. Stich, 
Phys. Rev. X {\bf 9}, 011018 (2019).

\bibitem{Li2015NatNano} 
L. Li, G. J. Ye, V. Tran, R. Fei, G. Chen, H. Wang, J. Wang, K. Watanabe, T. Taniguchi, L. Yang, X. H. Chen, and Y. Zhang, 
Nat. Nanotech. {\bf 10}, 608 (2015).

\bibitem{Gillgren20152DMat}
N. Gillgren, D. Wickramaratne, Y. Shi, T. Espiritu, J. Yang, J. Hu, J. Wei, X. Liu, 
Z. Mao, K. Watanabe, T. Taniguchi, M. Bockrath, Y. Barlas, R. K. Lake, and C. N. Lau, 
2D Mater. {\bf 2}, 011001 (2015).

\bibitem{Li2016NatNano}
L. Li, F. Yang, G. J. Ye, Z. Zhang, Z. Zhu, W. Lou, X. Zhou, L. Li, K. Watanabe, 
T. Taniguchi, K. Chang, Y. Wang, X. H. Chen and Y. Zhang, 
Nat. Nano. {\bf 11}, 593 (2016).

\bibitem{Zutic2004RMP}
I. \v{Z}uti\'c, J. Fabian, and S. Das Sarma,
Rev. Mod. Phys. {\bf 76}, 323 (2004).

\bibitem{Han2014NatNano}
W. Han, R. K. Kawakami, M. Gmitra, and J. Fabian, 
Nat. Nanotech. {\bf 9}, 794 (2014).

\bibitem{Kurpas2016PRB}
M. Kurpas, M. Gmitra, and J. Fabian, 
Phys. Rev. B {\bf 94}, 155423 (2016).

\bibitem{Avsar2017NatPhys}
A. Avsar, J. Y. Tan, M. Kurpas, M. Gmitra, K. Watanabe, T. Taniguchi, J. Fabian, and B. \"Ozyilmaz, 
Nat. Phys. {\bf 13}, 888 (2017).

\bibitem{Huber2017NatNano}
M. A. Huber, F. Mooshammer, M. Plankl, L. Viti, F. Sandner, L. Z. Kastner, T. Frank, J. Fabian, M. S. Vitiello, T. L. Cocker, and R. Huber, 
Nat. Nanotech {\bf 12}, 207 (2017).

\bibitem{Rodin2014PRL}
A. S. Rodin, A. Carvalho and A. H. Castro Neto, 
Phys. Rev. Lett. {\bf 112}, 176801 (2014).

\bibitem{Li2014PRB}
P. Li and I. Appelbaum, 
Phys. Rev. B {\bf 90}, 115439 (2014).

\bibitem{LewYanVoon2015NJP}
L. C. Lew~Yan~Voon, A. Lopez-Bezanilla, J. Wang, Y. Zhang, and M. Willatzen, 
New J. Phys. {\bf 17}, 025004 (2015).

\bibitem{PereiraJr2015PRB}
J. M. Pereira Jr. and M. I. Katsnelson, 
Phys. Rev. B {\bf 92}, 075437 (2015).

\bibitem{Kafaei2018JAP}
N. Kafaei, K. Beiranvand, M. Sabaeian, A. G. Dezfuli, and H. Zhang, 
J. Appl. Phys. {\bf 124}, 035702 (2018).

\bibitem{Kormanyos2013PRB}
A. Kormanyos, V. Zolyomi, N. D. Drummond, P. Rakyta, G. Burkard, and V. I. Fal'ko, 
Phys. Rev. B {\bf 88}, 045416 (2013).

\bibitem{Kormanyos2014PRX}
A. Kormanyos, V. Zolyomi, N. D. Drummond, and G. Burkard, 
Phys. Rev. X {\bf 4}, 011034 (2014).

\bibitem{Kormanyos2015NJP}
A. Kormanyos, P. Rakyta, and G. Burkard, 
New J. Phys. {\bf 17}, 103006 (2015).

\bibitem{Kane1957JPCS}
E. O. Kane, 
J. Phys. Chem. Solids {\bf 1}, 249 (1957).

\bibitem{Roth1959PR}
L. M. Roth, B. Lax, and S. Zwerdling, 
Phys. Rev. {\bf 114}, 90 (1959).

\bibitem{Hermann1977PRB}
C. Hermann and C. Weisbuch, 
Phys. Rev. B {\bf 15}, 823 (1977).

\bibitem{LewYanVoon2009}
L. C. Lew~Yan~Voon, M. Willatzen, 
The $k \cdot p$ method: electronic properties of semiconductors (Springer, Berlin, 2009).

\bibitem{Zhou2017PRB_gph}
X. Zhou, W.-K. Lou, D. Zhang, F. Cheng, G. Zhou and K. Chang, 
Phys. Rev. B {\bf 95}, 045408 (2017).

\bibitem{FariaJunior2016PRB}
P.~E. Faria~Junior, T. Campos, C.~M.~O. Bastos, M. Gmitra, J. Fabian, and G.~M. Sipahi, 
Phys. Rev. B {\bf 93}, 235204 (2016).

\bibitem{FariaJunior2019PRB}
P.~E. Faria~Junior, D. Tedeschi, M. De Luca, B. Scharf, A. Polimeni, and J. Fabian, 
Phys. Rev. B {\bf 99}, 195205 (2019).

\bibitem{Geim2013Nat}
A. K. Geim and I. V. Grigorieva, 
Nature {\bf 499}, 419 (2013).

\bibitem{Kormanyos20152DMat}
A. Kormányos, G. Burkard, M. Gmitra, J. Fabian, V. Zólyomi, N. D. Drummond, and V. Fal'ko, 
2D Materials {\bf 2}, 022001 (2015).

\bibitem{Zhou2017PRB_kpInSe}
M. Zhou, R. Zhang, J. Sun, W.-K. Lou, D. Zhang, W. Yang, and K. Chang, 
Phys. Rev. B {\bf 96}, 155430 (2017).

\bibitem{Brown1965}
A. Brown, S. Rundqvist, 
Acta Crystallographica {\bf 19}, 684 (1965)

\bibitem{QE2009}
P. Giannozzi, S. Baroni, N. Bonini, M. Calandra, R. Car, C. Cavazzoni, D. Ceresoli, G. L. Chiarotti, M. Cococcioni, I. Dabo et al., 
J. Phys. Cond. Mat. {\bf 39}, 395502 (2009);

\bibitem{QE2017}
P. Giannozzi, O. Andreussi, T. Brumme, O. Bunau, M. Buongiorno~Nardelli, M. Calandra, R. Car, C. Cavazzoni, D. Ceresoli, M. Cococcioni et al.,
J. Phys. Cond. Mat. {\bf 29}, 465901 (2017).

\bibitem{perdew_1996}
J. P. Perdew, K. Burke, and M. Ernzerhof, 
Phys. Rev. Lett. {\bf 77}, 3865 (1996); {\bf 78}, 1396(E) (1997).

\bibitem{wien2k}
P. Blaha, K. Schwarz, G. K. H.Madsen, D. Kvasnicka, and J. Luitz, 
{{WIEN2K}, {A}n {A}ugmented {P}lane {W}ave + {L}ocal {O}rbitals {P}rogram 
for {C}alculating {C}rystal {P}roperties}, {K}arlheinz Schwarz, Techn. Universit\"{a}t Wien, Austria, {2001}.

\bibitem{mBJ}
F. Tran and P. Blaha, 
Phys. Rev. Lett. {\bf 102}, 226401 (2009).

\bibitem{Dresselhaus1955PR}
G. Dresselhaus, 
Phys. Rev. {\bf 100}, 580 (1955).

\bibitem{Dresselhaus2008}
M. S. Dresselhaus, G. Dresselhaus and A. Jorio, 
{\it Group Theory: Application to the Physics of Condensed Matter}, 
1$^\textrm{st}$ Edition (Springer Berlin, 2008). 

\bibitem{supmat}
See Supplemental Material for details 
on the symmetry properties of phosphorene, the definitions of the $k \cdot p$ 
parameters, the explicit form of the different Hamiltonians and for details on 
the estimation of $\alpha_{c1v1}$ from the {\it ab initio} data.

\bibitem{lmfit}
M. Newville, T. Stensitzki, D. B. Allen and A. Ingargiola, 
{\it LMFIT: Non-Linear Least-Square Minimization and Curve-Fitting for Python¶}, 
(Zenodo , 2014), 10.5281/zenodo.11813.

\bibitem{Draxl2006CPC}
C. Ambrosch-Draxl and J. O. Sofo, 
Comp. Phys. Commun. {\bf 175}, 1 (2006).

\bibitem{Xu2016AdvMat}
R. Xu, J. Yang, Y. W. Myint, J. Pei, H. Yan, F. Wang, and Y. Lu, 
Adv. Mater. {\bf 28}, 3493 (2016).

\bibitem{Rudenko2015PRB}
A. N. Rudenko, Shengjun Yuan, and M. I. Katsnelson, 
Phys. Rev. B {\bf 92}, 085419 (2015).

\bibitem{FariaJunior2015PRB}
P.~E. Faria~Junior, G. Xu, J. Lee, N.~C. Gerhardt, G.~M. Sipahi, and I. \v{Z}uti\'c,
Phys. Rev. B {\bf 92}, 075311 (2015).

\bibitem{Chuang1997SST}
S. L. Chuang and C. S. Chang, 
Semicond. Sci. Technol. {\bf 12}, 252 (1997).

\bibitem{Long20162DMat}
G. Long, D. Maryenko, J. Y. Shen, S. G. Xu, J. Q. Hou, Z. F.
Wu, W. K. Wong, T. Y. Han, J. X. Z. Lin, Y. Cai, R. Lortz, and
N. Wang, 2D Mater. {\bf 3}, 031001 (2016).

\bibitem{Yang2018NL}
F. Yang, Z. Zhang, N. Z. Wang, G. J. Ye, W. Lou, X. Zhou, K. Watanabe, T. Taniguchi, K. Chang, X. H. Chen, and Y. Zhang, 
Nano Lett. {\bf 18}, 6611 (2018).

\bibitem{Pryor2006PRL}
C. E. Pryor and M. E. Flatt\'e, 
Phys. Rev. Lett. {\bf 96}, 026804 (2006).

\bibitem{vanBree2012PRB}
J. van~Bree, A. Yu. Silov, P. M. Koenraad, M. E. Flatt\'e, and C. E. Pryor, 
Phys. Rev. B {\bf 85}, 165323 (2012).

\bibitem{Zhou2015SREP}
X. Y. Zhou, R. Zhang, J. P. Sun, Y. L. Zou, D. Zhang, W. K. Lou, F. Cheng, G. H. Zhou, F. Zhai, and Kai Chang,
Scientific Reports {\bf 5}, 12295 (2015).

\bibitem{Tahir2015PRB}
M. Tahir, P. Vasilopoulos and F. M. Peeters, 
Phys. Rev. B {\bf 92}, 045420 (2015).

\bibitem{Rohlfing2000PRB}
M. Rohlfing and S. G. Louie, 
Phys. Rev. B {\bf 62}, 4927 (2000).

\bibitem{Scharf2016PRB}
B. Scharf, T. Frank, M. Gmitra, J. Fabian, I. \v{Z}uti\'c, and V. Perebeinos, 
Phys. Rev. B {\bf 94}, 245434 (2016).

\bibitem{Tedeschi2018PRBr}
D. Tedeschi, M. De Luca, P. E. Faria~Junior, A. Granados~del~\'Aguila, Q. Gao, 
H. H. Tan, B. Scharf, P. C. M. Christianen, C. Jagadish, J. Fabian and A. Polimeni, 
Phys. Rev. B {\bf 99}, 161204(R) (2019).

\bibitem{Rytova1967}
N. S. Rytova, Proc. MSU, Phys. Astron. {\bf 3}, 30 (1967).

\bibitem{Keldysh1979}
L. V. Keldysh, JETP Lett. {\bf 29}, 658 (1979).

\bibitem{Cudazzo2011PRB}
P. Cudazzo, I. V. Tokatly, and A. Rubio, 
Phys. Rev. B {\bf 84}, 085406 (2011).

\bibitem{Rodin2014PRB}
A. S. Rodin, A. Carvalho, and A. H. Castro Neto, 
Phys. Rev. B {\bf 90}, 075429 (2014).

\bibitem{Seixas2015PRB}
L. Seixas, A. S. Rodin, A. Carvalho, and A. H. Castro Neto, 
Phys. Rev. B {\bf 91}, 115437 (2015).

\bibitem{Prada2015PRB}
E. Prada, J. V. Alvarez, K. L. Narasimha-Acharya, F. J. Bailen, and J. J. Palacios, 
Phys. Rev. B {\bf 91}, 245421 (2015).

\bibitem{Berkelbach2013PRB}
T. C. Berkelbach, M. S. Hybertsen, and D. R. Reichman, 
Phys. Rev. B {\bf 88}, 045318 (2013).

\bibitem{Stier2018PRL}
A. V. Stier, N. P. Wilson, K. A. Velizhanin, J. Kono, X. Xu, and S. A. Crooker, 
Phys. Rev. Lett. {\bf 120}, 057405 (2018).

\bibitem{Haug2004}
H. Haug and S.~W. Koch,
{\it Quantum Theory of Optical and Electronic Properties of Semiconductors},
4$^\textrm{th}$ Edition (World Scientific Publishing, Singapore 2004).

\bibitem{Qiu2017NL}
D. Y. Qiu, F. H. da~Jornada and S. G. Louie,
Nano Lett. {\bf 17}, 4706 (2017).

\bibitem{Chaves2016PRB}
A. Chaves, M. Z. Mayers, F. M. Peeters, and D. R. Reichman, 
Phys. Rev. B {\bf 93}, 115314 (2016).

\bibitem{Tran2014PRB}
V. Tran, R. Soklaski, Y. Liang, and L. Yang, 
Phys. Rev. B {\bf 89}, 235319 (2014).

\bibitem{Tran20152DMat}
V. Tran, R. Fei, and L. Yang,
2D Mater. {\bf 2}, 044014 (2015).

\bibitem{Hunt2018PRB}
R. J. Hunt, M. Szyniszewski, G. I. Prayogo, R. Maezono, and N. D. Drummond, 
Phys. Rev. B {\bf 98}, 075122 (2018).

\bibitem{Li2019arXiv}
P. Li, arXiv:1908.10837 (2019).

\end{thebibliography}
\end{document}